\documentclass[12pt]{article}
\def\slashchar#1{\setbox0=\hbox{$#1$}           
   \dimen0=\wd0                                 
   \setbox1=\hbox{/} \dimen1=\wd1               
   \ifdim\dimen0>\dimen1                        
      \rlap{\hbox to \dimen0{\hfil/\hfil}}      
      #1                                        
   \else                                        
      \rlap{\hbox to \dimen1{\hfil$#1$\hfil}}   
      /                                         
   \fi}                                         %
\newcommand{\blank}{}
\renewcommand{\theequation}{\blank \arabic{equation}}
\newcounter{dummy}{}
\newcommand{\letters}{%
    \setcounter{dummy}{\value{equation}}
    \renewcommand{\thedummy}{\blank \arabic{dummy}}
    \renewcommand{\theequation}{\thedummy\alph{equation}}
    \refstepcounter{dummy}
    \setcounter{equation}{0}%
 }
\newcommand{\noletters}{%
    \setcounter{equation}{\value{dummy}}
    \renewcommand{\theequation}{\blank\arabic{equation}}%
  }
\newenvironment{mathletters}{\letters}{\noletters}
\newcommand{\half}{\frac 12}

\newcommand{\dirac}{\slashchar{\nabla}}
\newcommand{\Sferm}{\mathcal{S}}
\newcommand{\PP}{\mathcal{P}}

\newcommand{\ie}{\emph{i.~e.\/}}

\newcommand{\etc}{\emph{etc.\/\ }}
\begin{document}

\title{\bf The integral form of APS boundary conditions in the  Bag Model}
\author{A.~A.~Abrikosov,~jr.$^a$\footnote{{\bf e-mail}: persik@itep.ru},
A.~Wipf$\,^b$\footnote{{\bf e-mail}: wipf@tpi.uni-jena.de}
\\
$^a$ \small{\em Institute of Theoretical and Experimental Physics} \\
\small{\em B.~Cheremushkinskaya~25, Moscow 117~218, Russia}\\
$^b$ \small{\em  Institute of Theoretical Physics,
Friedrich-Schiller University} \\
\small{\em Max-Wien-Platz 1, 07743 Jena, Germany}
 }
\date{}
\maketitle

\begin{abstract}
We propose an integral form of Atiah-Patodi-Singer spectral
boundary conditions (SBC) and find explicitly the integral
projector onto SBC for the 3-dimensional spherical cavity. After
discussion of a simple example we argue that the relation
between the projector and fermion propagator is universal and
stays valid independently of the bag form and space dimension.
\end{abstract}

\section*{Introduction}

The two principal problems of QCD are confinement and spontaneous
breaking of chiral invariance. Both of them take place in the
strongly interacting domain where the theory becomes
nonperturbative. Most probably the two are interrelated. However,
usually they were considered separately. Up to now the spontaneous
chiral invariance breaking (SCIB) was discussed mostly in
infinite space. It would be interesting to study specific features
of SCIB that appear due to localization of quarks in finite
volume. In order to do that we confine the quarks in a
chiral invariant way.

One way to lock fermions in a finite volume without spoiling
chiral symmetry is to impose the so-called \textbf{spectral
boundary conditions} (SBC). They were introduced by Atiyah, Patodi
and Singer (APS) who investigated the spectral asymmetry for
manifolds with boundaries \cite{APS,euguchi/gilkey/hanson}. Later
these boundary conditions were widely applied in studies of
anomalies on manifolds with boundaries \cite{anomalies}.

Originally the APS boundary conditions were formulated for compact,
euclidean spin-manifolds in arbitrary even dimension $2n$. Such a manifold has a
compact $2n-1$, \ie\ odd-dimensional boundary. However in physical
applications one also meets odd-dimensional spatial bags evolving
in Euclidean or Minkowsky time. Evolution converts the spatial
boundary of such a bag into an infinite space-time cylinder.
Recently it was shown \cite{Abrikosov2006} that in these problems
instead of the full time-dependent boundary conditions on the
cylinder one may use static SBC on the spatial boundary of the
bag. Thus it turns out that the APS boundary conditions make sense
both on odd (the classical case) and even dimensional surfaces
(the modified SBC).

The spectral conditions are essentially nonlocal, \ie\ they are
defined on the boundary as a whole. According to the original
``constructive'' definition all spinor fields are expanded in terms of
eigenfunctions of the Dirac operator restricted to the boundary
and certain eigenfunctions in this expansion are required to be absent.
This ensures Hermiticity of the Dirac operator and charge
conservation in the bag.

However the SBC may also be written in an integral form.
Summation over separate boundary harmonics gives an integral
projection operator (we call it the SBC-projector).  The APS
conditions state that it must annihilate fermionic wave functions
on the boundary. The SBC-projector and its properties are
considered in the current paper.
We study the modified SBC for the
three-dimensional spherical bag. Its boundary is the Riemann
sphere $S^2$ and the corresponding harmonics may be expressed via
spherical spinors. In this case the summation may be carried out
and the SBC projection operator may be found explicitly.

A remarkable feature of our result is that the SBC projection
operator is immediately related to the fermionic Green function.
We claim that this relation is general and holds for arbitrary
non-spherical boundaries in any dimension.

The paper has the following structure. We shall review the
spectral boundary conditions in Section~1. In Section~2 we shall
discuss the APS conditions on the 3-dimensional sphere and
calculate the SBC-projector. Then, in Section~3 we shall discuss
the relation between the SBC-projector and fermion propagator first
for a simple model and then in the general case. These will be followed by
a summary and conclusions.

\section{The spectral boundary conditions}

\subsection{Conventions}

First we will introduce coordinates, Dirac matrices and fix
a gauge that allows us to formulate spectral boundary conditions.
The APS boundary conditions in any even dimension are defined
similarly and we may limit ourselves to the 4-dimensional case.
Generalization to higher dimensions is straightforward.

Let us consider massless fermions interacting with a gauge field
$\hat{A}$ in some Euclidean domain $B$ that may be either a closed
4-dimensional cavity $B_4$ or an infinite space-time cylinder
 $B_3 \otimes R$. We choose the curvilinear coordinates so that near the
boundary
 $\partial B$ the first coordinate $\xi$ points along the outward
normal while the three others, $q^i$, parametrize $\partial B$
itself. The origin $\xi=0$ lies on $\partial B$. Following the
classics we shall assume that near the surface the metric
 $g_{\alpha \beta}$ depends only on $q$ so that
\begin{equation}\label{g-APS}
  ds^2 =  d \xi^2 + g_{ik}(q)\, dq^i\, dq^k .
\end{equation}
For some subtleties concerning this point we refer to \cite{APSP}.
We choose the gauge such that on the boundary the normal
component
 $\hat{A}_\xi = 0$.

At this point it is convenient to choose the Dirac matrix $\gamma^\xi$. Let $I$ be the
$2\times 2$ unity matrix. Then
\begin{equation}\label{gamma-APS}
  \gamma^\xi = \left( \begin{array}{cc}
  0         & iI \\
  -iI & 0
\end{array} \right);
\qquad \qquad
  \gamma^q = \left( \begin{array}{cc}
  0         & \sigma^q \\
  \sigma^q  & 0
\end{array} \right),
\end{equation}
where $\{\sigma^1,\sigma^2,\sigma^2\}$ are the ordinary Pauli matrices. With
these definitions the Dirac operator for massless fermions on the surface
takes the form,
\begin{equation}\label{nabla-APS}
    \left.
        -i \slashchar{\nabla}
    \right|_{\partial B_4} =
         -i \gamma^\alpha \nabla_\alpha =
 \left(\begin{array}{cc}
  0                 & \hat{M}  \\
  \hat{M}^\dagger   & 0
\end{array} \right) =
 \left(\begin{array}{cc}
  0                 & I \, \partial_\xi - i \hat{\nabla}  \\
   - I \, \partial_\xi - i \hat{\nabla} & 0
\end{array} \right) ,
\end{equation}
where $\hat{\nabla} = \sigma^q\, \nabla_q$ is the convolution of
covariant gradient along the boundary $\nabla_q$ with
$\sigma$-matrices. Note that Hermitian conjugated operators
$\hat{M}$ and $\hat{M}^\dagger$ differ only by the sign
of~$\partial_\xi$.

Further on we shall call the linear differential operator
$-i \hat{\nabla}$ on the boundary the \textbf{boundary operator.}
 \begin{equation}\label{B-operator}
    \hat{B}_4 = -i \hat{\nabla} = -i \sigma^q\, \nabla_q.
\end{equation}
It is Hermitian and contains the tangential gauge field $\hat{A}_q$
and the spin connection that arises from the curvature of
$\partial B_4$.

It is well known that since the massless Dirac operator
anticommutes with $\gamma^5$-matrix,
\begin{equation}\label{chirality}
  \left \{-i \slashchar{\nabla},\, \gamma^5 \right\} = 0,
\qquad
        \gamma^5 = \left(
\begin{array}{rr}
  I & 0 \\
  0 & -I
\end{array} \right),
\end{equation}
it preserves helicity of massless quarks. This property is called
chiral invariance. In order to retain it in finite space one needs
chirally invariant boundary conditions.

\subsection{The APS boundary conditions} \label{APS}

\subsubsection{The classical 4-dimensional SBC}

Atiah, Patodi and Singer investigated spectra of Dirac operator
$-i\slashchar{\nabla}$ on manifolds with boundaries. If we
separate upper and lower (left and right) components of 4-spinors
the corresponding spectral equation will take the form
\begin{equation}\label{eigenvalue}
 -i\slashchar{\nabla}\, \psi_\Lambda =
 -i\slashchar{\nabla} \left(
\begin{array}{c}
u_\Lambda  \\ v_\Lambda
\end{array}\right) =
 \Lambda \left(
\begin{array}{c}
u_\Lambda  \\ v_\Lambda
\end{array}\right ) =
 \Lambda\, \psi_\Lambda.
\end{equation}
The next step in the construction of SBC is to Fourier-expand $u$
and $v$ near the boundary. Let 2-spinors $e_\lambda (q)$ be
eigenfunctions of the boundary operator $-i\hat{\nabla}$:
\begin{equation}\label{e-lambda}
   \hat{B}_4 \, e_\lambda (q) =
   -i\hat{\nabla}\, e_\lambda (q) =
   \lambda\, e_\lambda (q).
\end{equation}
Note that the form of this equation and the eigenfunctions
$e_\lambda (q)$ depend on the gauge-fixing. It is here that the gauge
condition
 $\hat{A}_\xi (0,\, q) = 0$ becomes important.

The operator $-i\hat{\nabla}$ is Hermitian so that its eigenvalues $\lambda$
are real. The functions $e_\lambda$ form an orthogonal basis on
 $\partial B_4$. In principle $-i\hat{\nabla}$ may have zero-modes
but for the sphere and convex manifolds this is not the case.

Due to assumption (\ref{g-APS}) in the vicinity of the boundary
the normal coordinate $\xi$ separates and the 2-spinors
$u_\Lambda$ and $v_\Lambda$ may be expanded in series:
\begin{mathletters}\label{uv-exp}
\begin{eqnarray}
  u_\Lambda (\xi,\, q) & = &
    \sum_\lambda f_\Lambda^\lambda (\xi)\, e_\lambda (q),
\quad
    f_\Lambda^\lambda (\xi)
        = \int_{\partial B_4}
        e_\lambda^\dagger (q)\, u_\Lambda (\xi,\, q)\,
        \sqrt g\, d^3 q;
\label{uv-exp/a} \\
  v_\Lambda (\xi,\, q) & = &
    \sum_\lambda g_\Lambda^\lambda (\xi)\, e_\lambda (q),
\quad
    g_\Lambda^\lambda (\xi)
        = \int_{\partial B_4}
        e_\lambda^\dagger (q)\, v_\Lambda (\xi,\, q)\,
        \sqrt g\, d^3 q;
\label{uv-exp/b}
\end{eqnarray}
\end{mathletters}
where $g=\det ||g_{ik}||$ is the determinant of the metric on the
boundary.

The APS boundary conditions may be defined in two equivalent ways:
either in terms of separate harmonics or via an integral operator.
\begin{itemize}
\item The traditional form of spectral boundary conditions states
      that on the boundary, \emph{i.~e.\/} at $\xi =0$
      \begin{mathletters}\label{SBC}
      \begin{eqnarray}
        \left.
          f_\Lambda^\lambda
        \right|_{\partial B_4} & = & 0
          \qquad \mathrm{for} \qquad \lambda > 0;  \label{SBC/a} \\
        \left.
          g_\Lambda^\lambda
        \right |_{\partial B_4} & = & 0
          \qquad \mathrm{for} \qquad \lambda < 0.  \label{SBC/b}
      \end{eqnarray}
      \end{mathletters}
    \item Another way is to introduce integral projectors
          $\mathcal{P}^+_4$ and $\mathcal{P}^-_4$ onto boundary modes with
          positive and negative $\lambda$ (4 means the 4-dimensional case):
          \begin{equation}\label{Projectors}
            \mathcal{P}^+_4 (q,\, q') =
              \sum_{\lambda > 0} e_\lambda (q) \, e_\lambda^\dagger (q');
          \qquad
            \mathcal{P}^-_4 (q,\, q') =
              \sum_{\lambda < 0} e_\lambda (q) \, e_\lambda^\dagger (q').
          \end{equation}
          If we join two-dimensional projectors $\mathcal{P}^+$
          and $\mathcal{P^-}$ into the $4\times 4$ integral
          operator $\mathcal{P}_4$ then the SBC for a 4-spinor $\psi$ will
          look as follows:
          \begin{equation}\label{P-APS}
          \mathcal{P}_4\, \psi (q) =
              \int_{\partial B_4}
              \left(
                  \begin{array}{cc}
                    \mathcal{P}^+_4 (q,\, q') & 0 \\
                    0 & \mathcal{P}^-_4 (q,\, q') \\
                  \end{array}
              \right)
              \left(
                  \begin{array}{c}
                    u (q') \\
                    v (q') \\
                  \end{array}
              \right)
              \sqrt g\, d^3 q' = 0.
          \end{equation}
\end{itemize}
The SBC-projector $\mathcal{P}_4$ commutes with $\gamma^5$,
and this boundary condition by construction respects chiral
invariance,
\begin{equation}\label{[P,gamma5]}
    \left[
        \mathcal{P}_4,\, \gamma^5
    \right] = 0.
\end{equation}
Let us denote by $\mathcal{I}$ the unity operator on the function
space spanned by the $e_\lambda$. Obviously, because of
completeness,
\begin{equation}\label{Ppls/Pmns}
    \mathcal{P}^+_4  (q,\, q') + \mathcal{P}^-_4 (q,\, q') =
    \mathcal{I}  (q,\, q') = I
        \left. \delta  (q - q')\right|_{\partial B_4},
\end{equation}
where the last expression is the $\delta$-function on the bag
surface $\partial B_4$.

\subsubsection{The truncated 3+1-dimensional SBC}

Now let us turn to fermions confined in a 3-dimensional spatial
bag $B_3$ that evolves in Euclidean time and sweeps the infinite
space-time cylinder $B_3\otimes R$. We will call the first three
coordinates ``space'' and the fourth one ``time''. The boundary
operator consists of spatial and temporal parts:
\begin{equation}\label{BOfull}
    \hat{B}_4 =
    -i \hat{\nabla}_{\partial B_3 \otimes R} =
    -i \hat{\nabla}_{\partial B_3} - i \sigma^z \partial_4 =
    \hat{B}_3 - i \sigma^z \partial_4.
\end{equation}
We will call the spatial part
 $\hat{B}_3 = -i \hat{\nabla}_{\partial B_3}$ the
\textbf{truncated boundary operator}. Let its eigenfunctions be
$e^\pm_\lambda$ (there was no ($\pm$)-superscript in
4-dimensions):
\begin{equation}\label{e-lambda-3d}
  -i\hat{\nabla}_{\partial B_3}\, e^\pm_\lambda (q) =
  \hat{B}_3\, e^\pm_\lambda (q) =
  \pm\lambda\, e^\pm_\lambda (q),
\qquad
    \lambda > 0.
\end{equation}

Wave functions on the space-time boundary $\partial B_3\otimes R$
can be expanded in $e^\pm_\lambda$ and longitudinal (temporal)
plane waves:
\begin{mathletters} \label{Fourier+/-}
\begin{eqnarray}
  u_\Lambda  & = &
    \sum_{\lambda >0}
    \int \frac{dk}{2\pi}\,e^{ikt}\,
    \left[
        f^{+\lambda,\, k}_\Lambda \, e^+_\lambda +
        f^{-\lambda,\, k}_\Lambda \, e^-_\lambda
    \right] ;
\label{Fourier+/-a}\\
  v_\Lambda  & = &
    \sum_{\lambda >0}
    \int \frac{dk}{2\pi}\,e^{ikt}\,
    \left[
        g^{+\lambda,\, k}_\Lambda \, e^+_\lambda +
        g^{-\lambda,\, k}_\Lambda \, e^-_\lambda
    \right] .
\label{Fourier+/-b}
\end{eqnarray}
\end{mathletters}

The truncated operator $-i \hat{\nabla}_{\partial B_3}$
anticommutes with $\sigma^z$ and because of that the temporal term
in (\ref{BOfull}) mixes positive and negative spatial harmonics.
This makes the full SBC: a)``future-dependent'' and b) hard to
handle.

However it was shown in \cite{Abrikosov2006} that in this case one
may use the simpler \textbf{truncated APS constraints}. In terms
of harmonics of the truncated boundary operator they look as
follows:
\begin{mathletters}\label{SBC3+1}
\begin{eqnarray}
    \left.
        f^{+\lambda,\, k}_\Lambda
    \right|_{\partial B_3} & = & 0 ;
\label{SBC3+1/a}
    \\
    \left.
        g^{-\lambda,\, k}_\Lambda
    \right|_{\partial B_3} & = & 0 .
\label{SBC3+1/b}
\end{eqnarray}
\end{mathletters}
These conditions are purely spatial and do not depend on time.
Thus they cause no problems with causality and may be applied both
in Euclidean and Minkowski spaces.

Projectors onto positive and negative boundary harmonics in the
(3+1)-case are defined in complete analogy with
(\ref{Projectors}):
\begin{equation}\label{3+1-Projectors}
  \mathcal{P}^\pm_{3+1} (q,\, q') =
  \sum_{\lambda > 0} e^\pm_\lambda (q) \, [e_\lambda^\pm
  (q')]^\dagger.
\end{equation}
This allows to put the truncated SBC in the integral form as
follows:
\begin{equation}\label{3+1-P-APS}
\mathcal{P}_{3+1}\, \psi (q) =
    \int_{\partial B_4}
    \left(
        \begin{array}{cc}
          \mathcal{P}^+_{3+1} (q,\, q') & 0 \\
          0 & \mathcal{P}^-_{3+1} (q,\, q') \\
        \end{array}
    \right)
    \left(
        \begin{array}{c}
          u (q') \\
          v (q') \\
        \end{array}
    \right)
    \sqrt g\, d^2 q' = 0,
\end{equation}
where the integration runs over the 2-dimensional \emph{spatial
boundary.}

\subsection{The physics of SBC}

It may be proved that SBC are chirally invariant, ensure Hermicity
of the Dirac operator and fermion conservation in the bag. The
interesting physical property of SBC is that fermionic wave
functions may be continued out of the bag in a square integrable
way.

Look how this may be shown. Let us write the eigenvalue equations
for Dirac operator near the boundary (remember that $\xi$ is the
outward spatial normal). In 3+1-di\-men\-sional case harmonics
corresponding to $\pm\lambda$ get mixed, therefore for each value
of $\lambda$ we get a set of four linked equations (instead of two
independent pairs for $k=0$ in $4d$):
\begin{mathletters}\label{RHS3+1}
\begin{eqnarray}
      (\partial_\xi
      + \lambda)\,
      g_\Lambda^{+\lambda,\, k}\,
      & = &
    \Lambda\,  f_\Lambda^{+\lambda,\, k}
    + ik\,  g_\Lambda^{-\lambda,\, k};
\label{RHS3+1/a}
\\
    - (\partial_\xi
     - \lambda)\, f_\Lambda^{+\lambda,\, k}
     & = &
     \Lambda\, g_\Lambda^{+\lambda,\, k}
    + ik\, f_\Lambda^{-\lambda,\, k} :
\label{RHS3+1/b}
\\
      (\partial_\xi
      - \lambda)\,
      g_\Lambda^{-\lambda,\, k}\,
      & = &
    \Lambda\,  f_\Lambda^{-\lambda,\, k}
    - ik\,  g_\Lambda^{+\lambda,\, k};
\label{RHS3+1/c}
\\
    - (\partial_\xi
     + \lambda)\, f_\Lambda^{-\lambda,\, k}
     & = &
     \Lambda\, g_\Lambda^{-\lambda,\, k}
    - ik\, f_\Lambda^{+\lambda,\, k} .
\label{RHS3+1/d}
\end{eqnarray}
\end{mathletters}
According to conditions (\ref{SBC3+1}) the RHS of equations
(\ref{RHS3+1/a}, \ref{RHS3+1/d}) vanish on the boundary. The
behaviour of $g^+$ and $f^-$ on the boundary is governed by
homogeneous equations so that:
\begin{equation}\label{logDf,g}
    \left.
        \frac{\partial_\xi f_\Lambda^{-\lambda,\, k} }%
        {f_\Lambda^{-\lambda,\, k}}
    \right|_{\xi=0} =
    \left.
        \frac{\partial_\xi g_\Lambda^{+\lambda,\, k} }%
        {g_\Lambda^{+\lambda,\, k}}
    \right|_{\xi=0} = -\lambda < 0.
\end{equation}

Thus the nonvanishing spinor components $g^+$ and $f^-$ have
negative logarithmic derivatives at the boundary. Therefore the
eigenfunctions may be continued exponentially out of both the
$4d$-bag or the $(3+1)d$-world cylinder in a square integrable
way. After the continuation the particles stay located mainly
inside the bag. This proves that SBC is a quite natural physical
requirement.

\section{The SBC-projector for sphere}

Now we are going to calculate explicitly the integral
SBC-projector for the boundary shaped as an ordinary 3-dimensional
Riemann sphere $S^2$. Properties of two-dimensional fermions on
the sphere were investigated in detail in \cite{Jayewardena}. The
eigenfunctions of Dirac operator in spherical coordinates were
found in \cite{DiracS2}. However, it is important that the form of
the boundary operator and, consequently, of SBC-projector varies
depending on the representation of spinors. A proper
representation may seriously reduce the complexity of the problem.
We deem our real success the implementation of what we call the
``work representation''. It notably simplifies the task and makes
the calculation feasible. This representation will be introduced
in the following section. Then we will outline the actual
computation and, finally, present the result in covariant form.

\subsection{Transformation of Dirac operator}

The form of the Dirac operator, boundary operator and SBC-projector
depend on choice of the $\gamma$-matrices. There exists a
coordinate-dependent transformation of spinors that converts the
eigenfunctions of the boundary operator on the sphere into
conventional spherical spinors. We call this the \textbf{work}
representation.

Let us start from standard Cartesian coordinates $x_\mu$, $\mu =
1,\dots\, 4$. The corresponding $\gamma$-matrices are:
\begin{equation}
 \label{gamma4d}
   \gamma_a = \left(
\begin{array}{rr}
  0 & \sigma_a \\
  \sigma_a & 0
\end{array}\right),
\quad
   \gamma_4 = \left(
\begin{array}{rr}
  0 & i I \\
  -iI & 0
\end{array}\right)
\quad
  \mbox{or}
\quad
    \gamma_\mu = \left(
\begin{array}{rr}
  0 & \sigma_\mu \\
  \sigma^\dagger_\mu & 0
\end{array}\right),
\end{equation}
with
 $\sigma_\mu= (\sigma_a,\, iI)$, $\sigma^\dagger_\mu= (\sigma_a,\,-iI)$.
The conventional rotation generators for 4-spinors are (for
spatial rotations it is convenient to use generators with one
index).
\begin{equation} \label{sigma-mu.nu}
  \Sigma_{\mu\nu} = -\frac i2 \left[ \gamma_\mu,\, \gamma_\nu \right]
\qquad
  \mbox{and}
\qquad
  \Sigma^a = \half \epsilon^{abc}\, \Sigma_{bc}.
\end{equation}
The Cartesian Dirac operator is:
\begin{equation}
\label{nabla3Cart}
  - i \dirac_{Cart} = - i \gamma_\mu \partial_\mu.
\end{equation}

Transformation to the work representation consists of two
independent $\frac \pi 2$-turns. The spatial turn about the radius
simplifies the boundary operator but leaves untouched the
 $\gamma_r = \gamma_a n_a$ matrix ($\vec n =\frac{\vec r}r$).
The rotation in $(r\, x_4)$-plane interchanges the $\gamma_4$ and
$\gamma_r$ matrices so that $\gamma_4$ points along the radius as
required by APS boundary conditions. The product of the two
rotations is:
\begin{equation}
\label{V4d-WF}
  V_W  =
   \exp \frac{i \pi}4\, \vec \Sigma\, \vec n \,
    \exp \frac{i \pi}4\, \Sigma_{a\, 4}\, n_a =
    \left(
    \begin{array}{cc}
    \frac{1 + i \vec \sigma \vec n}{\sqrt 2} & 0 \\
    0 & \frac{1 + i \vec \sigma \vec n}{\sqrt 2}
    \end{array}
    \right)
 \left(
    \begin{array}{cc}
    \frac{1 - i \vec \sigma \vec n}{\sqrt 2} & 0 \\
    0 & \frac{1 + i \vec \sigma \vec n}{\sqrt 2}
    \end{array}
    \right) =
     \left(
    \begin{array}{cc}
    I & 0 \\
    0 & i \vec \sigma \vec n
    \end{array}
    \right);
\end{equation}
The result of the transformation on the matrices $\gamma_4$ and
$\gamma_r$ is:
\begin{equation}
\label{WF.gamma}
  V_W^\dagger\, \vec \gamma \vec n\, V_W  = \gamma_4;
  \qquad
  V_W^\dagger\,  \gamma_4\, V_W  = - \vec \gamma \vec n;
\end{equation}
After rotation to the work frame the Dirac operator takes the
form:
\begin{equation}
  \label{DW}
  -i  \dirac_W =
  - i V_W^\dagger\, \dirac_{Cart} \, V_W =
\left(
  \begin{array}{cc}
    0 &
    \left(
      \partial_r+\frac 1r
    \right)
     +\frac {\hat B_W}r \\
     - \left(
      \partial_r+\frac 1r
    \right)
     +\frac {\hat B_W}r & 0
  \end{array}
\right)
  + i\, \vec \gamma \vec n\, \partial_4,
\end{equation}

Up to the term $\frac 1r$ that accompanies the $\partial_r$-derivative (it
may be eliminated by redefining the wave function
 $\psi \rightarrow \psi/r$) this operator has exactly the required
form ($\gamma_4$ is aligned with the radius). Therefore we may
project solutions of the Dirac equation onto eigenfunctions of
 $\hat B_W$ and impose on them the modified APS-boundary conditions.

\subsection{The boundary operator}

For convenience let us introduce (3+1) cylindrical coordinates:
\begin{equation} \label{3+1cyl}
 x^1=r\sin \theta \cos \phi ;   \quad
 x^2=r\sin \theta \sin \phi ;   \quad
 x^3=r\cos \theta ;     \quad
 x^4=t.
\end{equation}
After the spinor rotation (\ref{WF.gamma}) the boundary operator
in (\ref{DW}) takes the following form:
\begin{equation}
\label{B3work}
  {\hat B_W} =
  \left(
  \begin{array}{cc}
    -i \partial_\phi +1 &
    - e^{- i \phi}
    (\partial_\theta -i \cot\theta\,\partial_\phi)
    \\
     e^{i \phi}
    (\partial_\theta + i \cot\theta\,\partial_\phi) &
     i \partial_\phi +1
  \end{array}
\right) =
  I + 2 \hat{L}^a_W \hat{S}^a_W.
\end{equation}
Thus eigenfunctions of the boundary operator are classified
according to the value of scalar product
 $\hat{L}^a_W \hat{S}^a_W$. The operator $\hat{S}^a_W$
looks like the nonrelativistic spin,
\begin{equation}\label{SW-operator}
     \hat{S}^a_W = \half \sigma^a,
\end{equation}
and $\hat{\vec L}_W$ is the 3-dimensional angular momentum in
spherical coordinates (in Cartesian frame it would be
 $\hat L^a_W = \hat{L}^a = -i\epsilon^{abc} x_b \partial_c$).

Note, that despite the apparent resemblance $\hat{\vec S}_W$ is
not the physical spin. Rotation~(\ref{WF.gamma}) affects only the
lower (right) components of 4-spinors and makes the actual spin
operators
for left and right fields look differently%
\footnote{%
The true $4d$ spin operator in Cartesian frame is $\half
\vec \Sigma$. After the rotation to work representation it
becomes:
\begin{equation}\label{Spin-W}
    \half \vec \Sigma_W =
    \half V_W^\dagger\, \vec \Sigma\, V_W  =
    \half \left(
        \begin{array}{rr}
          \vec \sigma & 0 \\
          0 & 2 \vec n (\vec \sigma\, \vec n) - \vec \sigma
        \end{array}
    \right)
\end{equation}%
}.

However, one still may profit from the formal similarity of $S_W$
to spin. Let us introduce a fictitious operator of ``total angular
momentum'' $\hat{J}_W = \hat{L}_W + \hat{S}_W,$. Then the boundary
operator in the work representation (\ref{B3work}) may be written
as
\begin{equation}\label{B3W(JL)}
    \hat B_W = I + 2 \hat{L}^a_W \hat{S}^a_W =
    I + \hat J_W^2 - \hat L_W^2 - \hat S_W^2.
\end{equation}

This beautiful formula is one of our main results and the key
element of the further calculation. The reduction of spherical
Dirac operator to momentum operators greatly simplifies all
spectral expansions including that of the SBC-projector. Obviously
the eigenfunctions of the operator (\ref{B3W(JL)}) are
conventional spherical spinors $\Omega_{j,\, l,\, m}$ and the
corresponding eigenvalues of $\hat B_W$ are (see
eqn.~(\ref{e-lambda-3d})),
\begin{equation}\label{B3W-values}
    \lambda =
    j(j+1) -l(l+1) +\frac14 =
    \left \{
    \begin{array}{rcc}
      l+1, \quad &\mathrm{for} &\quad  j = l + \half ; \\
      -l, \quad  &\mathrm{for} &\quad  j = l - \half . \\
    \end{array}
    \right.
\end{equation}
Explicit expressions for spherical spinors will be given in the
next section.

\subsection{Calculation of the projector}

Now we are going to calculate the 3-dimensional projectors
 $\PP^\pm_{3+1} (q,\, q')$ defined by
formulae~(\ref{3+1-Projectors}). The eigenfunctions of the
boundary operator $\hat B_W$ are the standard spherical spinors.
In terms of the conventional spherical harmonics $Y_{l,\, m}$ they
look as follows:
\begin{equation}\label{Ojlm}
  e^+_{l+1} =
  \Omega_{l+\half,\, l,\, k} = \left(
\begin{array}{r}
  \sqrt{\frac{j+k}{2j}} \, Y_{l,\, k-\half} \\
  \sqrt{\frac{j-k}{2j}} \, Y_{l,\, k+\half}
\end{array}\right)
    \quad \mathrm{and} \quad
  e^-_{l} =
  \Omega_{l-\half,\, l,\, k} = \left(
\begin{array}{r}
  - \sqrt{\frac{j-k+1}{2j+2}} \, Y_{l,\, k-\half} \\
  \sqrt{\frac{j+k+1}{2j+2}} \, Y_{l,\, k+\half}
\end{array}\right).
\end{equation}
Substituting them into equations~(\ref{3+1-Projectors}) we get for
the projectors in the work representation (here $m=k-\half$ and
 $c_{lm} = \sqrt{(l-m)(l+m+1)}$):
\begin{mathletters}\label{Sum+/-}
    \begin{eqnarray}
      \lefteqn{\PP^+_W (\hat x,\, \hat y) =
      \sum^\infty_{l=0}\sum_{m=-l-1}^l
      \frac{1}{2l+1} \times}
      \hspace{10mm} & &
      \nonumber \\
      & &
    \left(
    \begin{array}{cc}
     (l+m+1)\, Y_{l,\, m}(\hat x)\, Y^*_{l,\, m}(\hat y) &
     c_{lm}\,
     Y_{l,\, m}(\hat x)\, Y^*_{l,\, m+1}(\hat y)
     \\
     c_{lm}\,
     Y_{l,\, m+1}(\hat x)\, Y^*_{l,\, m}(\hat y) &
    (l-m)\, Y_{l,\, m+1}(\hat x)\, Y^*_{l,\, m+1}(\hat y) \\
    \end{array}
    \right ) ;
    \label{Sum+/-/a} \\
      \lefteqn{\PP^-_W (\hat x,\, \hat y) =
      \sum^\infty_{l=1}\sum_{m=-l}^{l-1}
      \frac{1}{2l+1} \times}
      \hspace{10mm} & &
      \nonumber \\
      & &
    \left(
    \begin{array}{cc}
     (l-m)\, Y_{l,\, m}(\hat x)\, Y^*_{l,\, m}(\hat y) &
     - c_{lm}\,
     Y_{l,\, m}(\hat x)\, Y^*_{l,\, m+1}(\hat y)
     \\
     - c_{lm}\,
     Y_{l,\, m+1}(\hat x)\, Y^*_{l,\, m}(\hat y) &
    (l+m+1)\, Y_{l,\, m+1}(\hat x)\, Y^*_{l,\, m+1}(\hat y) \\
    \end{array}
    \right ) .
      \label{Sum+/-/b}
    \end{eqnarray}
\end{mathletters}
The summation is a purely technical problem. At the end we get for
the full 4-component projector (\ref{3+1-P-APS}):
\begin{equation}\label{P-S-Work}
    \left.
    \PP_W (\hat x,\, \hat y)
    \right|_{\hat x,\, \hat y \in S^2}  =
    \half \delta_{S^2} (\hat x - \hat y) -
    i\, \gamma^4\, \Sferm_W (\hat x,\, \hat y),
\end{equation}
where $\Sferm_W$ is the propagator of massless fermions in the
work representation.

After rotating back to the Weyl representation (\ref{gamma4d}) and
substituting the fermion propagator $\Sferm$ we get the explicit
form of APS boundary condition on a sphere of radius $R$:
\begin{eqnarray}\label{APS-cov}
    \lefteqn{
    \left.
    \PP\, \psi(\vec x)\right|_{|x| = R} =
    \oint_{S^2}
    \left[
    \half \delta_{S^2} (\vec x - \vec y) +
    \frac{\slashchar{\hat n}_x\,
     (\slashchar{\vec x} - \slashchar{\vec y})}%
    {4\pi\, |\vec x - \vec y|^3}
    \right]\,
    \psi(\vec y)\, d^2 y =} & &
    \nonumber \\
    & & \hspace{15mm}
    \oint_{S^2}
    \left[
    \half \delta_{S^2} (\vec x - \vec y) +
    \frac{(R^2 -\slashchar{\vec x} \slashchar{\vec y})}%
    {4\pi\,R\, |\vec x - \vec y|^3}
    \right]\,
    \psi(\vec y)\, d^2 y = 0
\end{eqnarray}
This completes the calculation in 3-dimensional case.
Now we are going to illustrate the mechanism that causes the
appearance of the fermionic Green function by an example.

\section{The two-mode model}\label{2mode}

Here we will demonstrate how the relation between the
SBC-projector and fermion propagator arises at the level of a
separated boundary harmonic. This may be done by studying the
simple model with an almost trivial boundary operator.

Let us consider two-dimensional massless fermions living in
half-plane $\xi < 0$. Suppose that the boundary operator has a
single eigenvalue $\lambda > 0$ (actually this means that we
consider an individual mode of the boundary operator). Then the
``Dirac operator'' $\hat{\Lambda}$ is:
\begin{equation}\label{2mode/Dirac}
    \hat{\Lambda} =
    - i \gamma^\xi  \partial_\xi  + \hat{\lambda} =
    \left(
          \begin{array}{cc}
            0 & \partial_\xi \\
            -\partial_\xi & 0
          \end{array}
    \right) +
    \left(
          \begin{array}{cc}
            0 & \lambda \\
            \lambda & 0
          \end{array}
    \right).
\end{equation}
The Dirac equation $\hat{\Lambda}\psi = 0$ has two solutions:
\begin{equation}\label{2mode/solutions}
    \psi (\xi ) =
    c^-\, \psi^- (\xi) + c^+\, \psi^+ (\xi) =
    c^-\, \exp(- \lambda \xi)
    \left(
          \begin{array}{c}
            0 \\
            1
          \end{array}
    \right) +
    c^+\, \exp(\lambda \xi)
    \left(
          \begin{array}{c}
            1 \\
            0
          \end{array}
    \right).
\end{equation}
The projectors onto positive and negative modes and unity operator are
simple $2\times 2$ matrices:
\begin{equation}\label{2mode/projectors}
    P^+ =
    \left(
          \begin{array}{cc}
            1 & 0 \\
            0 & 0
          \end{array}
    \right);
\qquad
    P^- =
    \left(
          \begin{array}{cc}
            0 & 0 \\
            0 & 1
          \end{array}
    \right);
\qquad
    I = P^+ + P^- =
    \left(
          \begin{array}{cc}
            1 & 0 \\
            0 & 1
          \end{array}
    \right).
\end{equation}
The APS boundary condition in this case kills the positive mode
that grows at $+\infty$ and has the form
\begin{equation}\label{2mode/APS}
    P\, \psi (\xi ) =
    P^+\,\psi (\xi ) =
    \left(
          \begin{array}{cc}
            1 & 0 \\
            0 & 0
          \end{array}
    \right)
    \psi (\xi ) = 0.
\end{equation}
Now let us turn to the Green function of operator
(\ref{2mode/Dirac}). It is easy to check that
\begin{equation}\label{2mode/propagator}
    \hat{\Lambda}^{-1} = S (\xi,\, \eta) =
    \left(
          \begin{array}{cc}
            0 & \theta (\eta - \xi)\, e^{\lambda(\xi - \eta)}  \\
             \theta (\xi - \eta)\, e^{\lambda(\eta - \xi)} & 0
          \end{array}
    \right);
\qquad
    \hat{\Lambda} \,S (\xi,\, \eta) =
    I\, \delta (\xi - \eta).
\end{equation}
In order to establish a connection between the fermion propagator
and SBC-projector we put both points $\xi, \, \eta$ onto the
boundary, $\xi = \eta= 0$, as it was done in (\ref{P-S-Work}), and
multiply $S$ by $-i \gamma^\xi$:
\begin{equation}\label{2mode/S(x=y)}
    \left.
          -i \gamma^\xi S (\xi,\, \eta)
    \right|_{\xi = \eta= 0} =
    \left(
          \begin{array}{cc}
            \theta (0) & 0  \\
             0 & - \theta (0)
          \end{array}
    \right) =
    \theta (0) (P^+ - P^-)  =
    \frac{P^+ - P^-}2 .
\end{equation}
Here we used the symmetric regularization and set
 $\theta (0)=\half$. Comparison with (\ref{2mode/projectors})
demonstrates that the projector onto the APS boundary conditions
(\ref{2mode/APS}) may be expressed as
\begin{equation}\label{2mode/P-S}
    P^+ =
    \half \, I
      -i \left.
            \gamma^\xi S (\xi,\, \eta)
      \right|_{\xi = \eta= 0} ,
\end{equation}
that is exactly the one-dimensional version of
Eq.~(\ref{P-S-Work}).

We conclude that the result for the model is similar to that obtained
by an exact calculation in the more realistic case. Actually the
situation is quite general and the relation between the SBC-projector
and fermionic Green function stays valid regardless of
dimensionality of space, shape of the boundary and spinor
representation. Indeed, the only simplification of our example is
that the boundary operator had a single eigenvalue. Although the real
spectra of boundary operators are much richer, our analysis remains
true for every single isolated eigenfunction. Being valid for each
of the spectral harmonics it must stay true for the functions as a
whole.

\section{Conclusion}\label{conclusion}

Here is the brief summary of our results. We proposed the integral
form of SBC for static even-dimensional bags and truncated SBC for
odd-dimensional bags evolving in time. Then we performed the
explicit calculation of SBC-projector for the 3-dimensional
sphere. The projector has a simple and compact expression in terms
of the massless fermion propagator. The analysis of a simple
two-mode model of APS boundary conditions led to the same
relation between the projector and propagator.

The covariant form of our result may be obtained by transforming
(\ref{P-S-Work}) to Weyl representation.  Then the projector onto
APS boundary conditions becomes:
\begin{equation}\label{P-S-Cart}
    \PP (\hat x,\, \hat y) =
    V_W\, \PP_W (\hat x,\, \hat y)\, V_W^\dagger =
    \half \delta_{\partial B} (\hat x - \hat y) -
    i\, (\hat \gamma\, \hat n_x)\, \Sferm (\hat x,\, \hat y),
\end{equation}
where vectors $\hat x,\, \hat y \in \partial B$ and $\hat n_x$ is
the outward normal to the boundary at point $\hat x$. Here
$\Sferm$ is the conventional propagator of massless fermions. This
expression does not depend on parametrization of $\gamma$-matrices
and choice of coordinates.

The relation between the SBC-projector and massless fermion
propagator (\ref{P-S-Cart}) is covariant and does not depend on
representation of spinors. Therefore the integral formulation of
spectral boundary conditions looks more practical than the
original ``constructive'' definition that required the special
choice of coordinates $\gamma$-matrices \etc Moreover, the result
(\ref{P-S-Cart}) appears to be universal and suitable for
boundaries of arbitrary shape in any dimension. It is not
difficult to prove this in general starting from the two-mode
example of Section~\ref{2mode}.

\section{Acknowledgements}
In conclusion we are glad to thank D.~Laenge, A.~Kirchberg,
K.~Kirsten and M.~Santangelo for discussions on various aspects of
APS boundary conditions. A.~Abrikosov expresses his gratitude for
the DAAD sponsorship that made possible the present collaboration.
His work was also partially supported by RFBR grants
\mbox{05--02--17464} and \mbox{06--02--16905}.


\end{document}